\documentclass[journal]{IEEEtran}
\ifCLASSINFOpdf
\else
\fi

\usepackage{graphicx}
\usepackage{bm}
\usepackage{amsfonts}
\usepackage{amsmath}
\usepackage{amssymb}
\usepackage{geometry}
\usepackage{times}
\usepackage{cite}
\usepackage{subfigure}
\usepackage{latexsym,bm,amsmath,amssymb} 
\usepackage{CJK}
\usepackage{hhline}

\usepackage{multirow} 
\usepackage{amsmath}
\usepackage{xcolor}
\usepackage[linesnumbered,ruled,vlined]{algorithm2e}
\usepackage{url}
\newcommand{\tb}{\mathbf}
\newcommand{\revise}{\textcolor{black}}

\begin{document}

%
\title{Orthogonal Time Frequency Space Modulation -- Part III: ISAC and Potential Applications}
\author{Weijie Yuan,~\IEEEmembership{Member,~IEEE}, Zhiqiang Wei,~\IEEEmembership{Member,~IEEE}, Shuangyang Li,~\IEEEmembership{Member,~IEEE}, \\Robert Schober,~\IEEEmembership{Fellow,~IEEE}, and Giuseppe Caire,~\IEEEmembership{Fellow,~IEEE}\\
(\emph{Invited Paper})
\vspace{-8mm}
\thanks{
W. Yuan is with the Department of Electrical and Electronic Engineering, Southern University of Science and Technology, Shenzhen 518055, China (e-mail: yuanwj@sustech.edu.cn).

 Z. Wei is with the School of Mathematics and Statistics, Xi'an Jiaotong University, Xi'an 710049, China (e-mail: zhiqiang.wei@xjtu.edu.au).
 
 S.~Li was with the School of Electrical Engineering and Telecommunications, University of New South Wales, Sydney, NSW 2052, Australia, when this letter was submitted. He is now with the Department of Electrical, Electronic, and Computer Engineering, University of Western Australia, Perth, WA 6009, Australia (e-mail: shuangyang.li@uwa.edu.au).
 
 R. Schober is with the Institute for Digital Communications (IDC), the Friedrich-Alexander University Erlangen-Nuremberg, Erlangen 91054, Germany (e-mail: robert.schober@fau.de).

 G. Caire is with the Department of Electrical Engineering and Computer Science, Technical University of Berlin, Berlin 10587, Germany (caire@tu-berlin.de).
}
}



\maketitle


\begin{abstract}
The first two parts of this tutorial on orthogonal time frequency space (OTFS) modulation have discussed the fundamentals of delay-Doppler (DD) domain communications as well as some advanced technologies for transceiver design. In this letter, we will present an OTFS-based integrated sensing and communications (ISAC) system, which is regarded as an enabling technology in next generation wireless communications. In particular, we illustrate the sensing as well as the communication models for OTFS-ISAC systems. Next, we show that benefiting from time-invariant DD channels, the sensing parameters can be used for inferring the communication channels, leading to an efficient transmission scheme. As both functionalities are realized in the same DD domain, we briefly discuss several promising benefits of OTFS-based ISAC systems, which have not been completely unveiled yet. Finally, a range of potential applications of OTFS for the future wireless networks will be highlighted.

\end{abstract}


\IEEEpeerreviewmaketitle
\section{Introduction}
The orthogonal time frequency space (OTFS) modulation has gained considerable attentions due to its capability of providing more reliable communications compared to the existing orthogonal frequency division multiplexing (OFDM) modulation, especially in high-mobility environments \cite{weiotfs}. As overviewed in Part I of this tutorial, OTFS utilizes the delay-Doppler (DD) domain for data multiplexing, which mirrors the geometry of the scatterings comprising the wireless channel. In history, the delay and Doppler parameters are typically used for sensing-related applications, representing the range and velocity characteristics corresponding to the moving targets. In future wireless networks, both sensing and communications functionalities are highly desired, leading to the integrated sensing and communications (ISAC) technology by using the same signaling waveform and hardware architecture \cite{saddik2007ultra,liu2020joint}. In contrast to the OFDM-based ISAC system, OTFS-ISAC system provides the direct interaction between transmitted signals and the channel responses in a unified DD domain for both functionalities, on top of the resilience to delay and Doppler spreads. To this end, OTFS has been regarded as a promising signal waveform to fully unleash the potentials of ISAC system.

The pioneering work \cite{raviteja2019orthogonal} compared the radar sensing performance based on OTFS and OFDM schemes, showing that OTFS is more robust for velocity estimation. Some recent works discussed the ISAC performance limits, the beamforming designs, the resource allocation schemes, and the receiver structures for OTFS-based ISAC systems \cite{9109735,mariotfs,9557830,9724198,wu2021otfs}. All these contributions have demonstrated the effectiveness of OTFS for ISAC systems. However, considering the inherent relationship between DD domain communications and sensing, there are several promising benefits of adopting OTFS as the waveform for realizing ISAC, which have been not well discussed in the literature.

In this letter, we will first study the OTFS signaling model for both communications and sensing in the DD domain. It is noted that the delay and Doppler associated with the DD domian communication channel are half of the round-trip delay and Doppler parameters extracted from the sensing echoes. This motivates us that the sensing parameters can be used for predicting the communication channels given the slow time-varying property of the DD domain channel. Afterwards, we briefly discuss several benefits of OTFS-ISAC system, i.e., unified signaling design framework, sensing channel exploitation, and communications-assisted sensing. Finally, we discuss potential applications related to OTFS modulation, including underwater communications, optical transmission, and space-air-ground integrated networks.

\emph{Notations:}$\left(\cdot\right)^{\rm H}$ and $\left(\cdot\right)^{\rm T}$ denotes the Hermittian operation and the transpose operation, respectively; $\left(\cdot\right)_N$ denotes the modulus operation w.r.t. $N$; $\tb{I}_{MN}$ denotes an identity matrix of dimension $MN$; $\delta\left(\cdot\right)$ denotes the Dirac delta function; $\mathbb{E}[\cdot]$, $\textrm{Tr}[\cdot]$, and $\textrm{det}[\cdot]$ denote the expectation operation, trace operation, and the determinant of a matrix, respectively.

\section{OTFS-ISAC System Model}
\begin{figure}[!t]
\centering
\includegraphics[width=.4\textwidth]{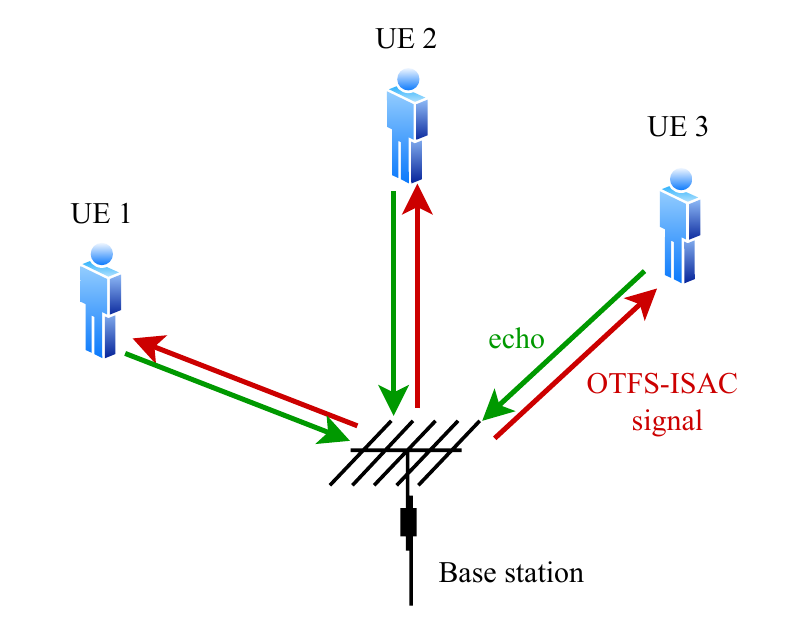}
\caption{{OTFS-ISAC system model.}}%
\label{model}%
\vspace{-5mm}
\end{figure}

Let us consider a general system model with a mono-static base station (BS) and $P$ targets of interests, as shown in Fig. \ref{model}. For ease of exposition, we assume that all targets of interest are communications user equipments (UEs) as well, which aligns well with the use case such as vehicular networks. The BS is equipped with a transmit uniform linear array (ULA) of $N_t$ antennas and a separate receive ULA of $N_r$ antennas. Under the assumption of sufficient isolation between the transmit and receive arrays, the sensing echoes would not interfere with the transmitted signals. For receiving the communication information, each target is equipped with $N_u$ antennas.

\subsection{Communication Model}
Without loss of generality, we consider a DD domain grid of $M\times N$, where $l \in \{0,\ldots,M-1\}$ and $k \in \{0,\ldots,N-1\}$ denotes the delay and Doppler indices, respectively. The information symbols sent to the $i$-th target are defined as $X_{\rm DD}^{i}\left[l,k\right]$, which is transformed to the time-frequency (TF) domain via inverse symplectic
finite Fourier transform (ISFFT), yielding $X^i_{\rm TF}\left[m,n\right]$, where $m$ and $n$ denote the indices of subcarriers and time slots. By performing multicarrier modulation, the transmitted signal to the $i$-th target is given by
\begin{align}\label{tftot}
\hspace{-2mm}s_i(t)=\sum_{n=0}^{N-1}\sum_{m=0}^{M-1}X^i_{\rm TF}\left[m,n\right] g_{\rm tx}(t-nT) e^{j2\pi m \Delta f (t-nT)} ,
\end{align}
where $g_{\rm tx}(t)$ is the transmit shaping pulse, $T$ and $\Delta f$ denote the symbol duration and frequency spacing, respectively, satisfying $T\Delta f = 1$ to maintain the orthogonality.

Initially at the target detection stage, the BS could use the antenna array to formulate a wide beam or an omni-directional signal to detect all targets of interests. Afterwards, in the tracking mode, the BS formulates multiple beams for information transmission and target tracking. Through a beamforming matrix $\tb{F}\in\mathbb{C}^{N_t\times P}$, the multi-beam signal is expressed as
\begin{align}
\tilde{\tb{s}}(t) = \tb{F} \tb{s}(t),
\end{align}
where $\tb{s}(t) = \left[s_1(t),...,s_P(t)\right]^{\rm T}$ is the transmitted signal to all $P$ targets.
\revise{The $i$-th column of $\tb{F}$, denoted by $\tb{f}_{i} = \sqrt{\frac{p_i}{N_t}}\tb{a}_{N_t}(\tilde{\theta}_i)$ is used to allocate power $p_i$ and to steer the transmitted signal towards the intended direction $\tilde{\theta}_i$},
where the steering vector is given by
\begin{align}
\tb{a}_{N_t}(\theta_i) = \left[1,e^{j \pi \sin\theta_i},...,e^{j (N_t-1)\pi \sin\theta_i}\right]^{\rm T}.
\end{align}

\revise{Due to the asymptotic orthogonality of the massive antenna array, after transmit beamforming, the communication channel to the $i$-th target is asymptotically line-of-sight (LoS) dominated}, and given by
 \begin{align}
 	\tb{C}^i_{\rm DD}(\tau,\nu) = h_i \tb{a}_{N_u}(\theta_i)  \tb{a}_{N_t}^{\rm H}(\theta_i)  \delta(\tau-\tau_i) \delta(\nu-\nu_i),
 \end{align}
 with channel gain $h_{i}= \sqrt{\frac{c}{4\pi f_c d^2_i}}$, signal propagation speed $c$, carrier frequency $f_c$, range distance $d_i$, delay $\tau_i$, and Doppler shift $\nu_i$. After passing through the channel \cite{9082873}, the received signal of the $i$-th target can be modelled as
 \begin{align}
 {y}_i(t) = h_{i} \tb{u}^{\rm H}_i \tb{a}_{N_u}(\theta_i)  \tb{a}_{N_t}^{\rm H}(\theta_i) \tb{f}_i s_i(t-\tau_i) e^{j 2 \pi \nu_i (t-\tau_i)} + z_i(t),	
 \end{align}
where $\tb{u}_i\in\mathbb{C}^{N_u \times 1}$ is the receive beamformer, $z_i(t)$ is the time domain noise signal. We adopt a receive filter $g_{\rm rx}(t)$ and after OTFS demodulation, the received OTFS signal is expressed as\footnote{\revise{For simplicity, we assume integer delay and Doppler shifts in this letter. For the fractional case, off-grid estimation schemes can be adopted, e.g., \cite{wei2022off}.}}
\begin{align}\label{downlink_receive}
	Y^i_{\rm DD}\left[ {l,k} \right]  =& h_i \tb{u}^{\rm H}_i \tb{a}_{N_u}(\theta_i)  \tb{a}_{N_t}^{\rm H}(\theta_i) \tb{f}_i X^i_{\rm DD}\left[ {(l-l_i)_M,(k-k_i)_N} \right] \nonumber\\&+Z^i_{\rm DD}\left[ {l,k} \right],
\end{align}
where the integers $k_i = {\nu_i}{N T}$, $l_i = {\tau_i}{M \Delta f}$, $Z^i_{\rm DD}\left[ {l,k} \right]$ is the additional white Gaussian noise sample with a power spectral density of $N_0$.

\subsection{Sensing Model}

The sensing channel is given by
\begin{align}
\tb{H}(t,\tau)  = \sum_{i=1}^K \gamma_i \tb{a}_{N_r}(\theta_i) \tb{a}_{N_t}^{\rm H}(\theta_i) \delta(\tau-\eta_i) e^{j2\pi \upsilon_i t},
\end{align}
where $\gamma_i$, $\eta_i$, and $\upsilon_i$ are the reflected coefficient, the delay, and the Doppler associated with the $i$-th target. Then, the acquired sensing echoes at the BS can be expressed as
\begin{align}
\tb{r}(t) = \sum_{i=1}^K \gamma_i \tb{a}_{N_r}(\theta_i) \tb{a}_{N_t}^{\rm H}(\theta_i) \tilde{\tb{s}}(t-\eta_i)e^{j2\pi \upsilon_i t} + \tb{w}(t),
\end{align}		
where $\tb{w}(t)$ is the measurement noise. \revise{For massive MIMO receive antenna arrays, steering vectors with different angular values are asymptotically orthogonal \cite{marzetta2016fundamentals}, i.e., $\tb{a}_{N_r}^{\rm H}(\theta_i)\tb{a}_{N_r}(\theta_{i'})\approx 0,~\forall \theta_i\neq \theta_{i'}$. Thus, the interference between different targets in the sensing echoes can be neglected and the BS can distinguish different targets in terms of their angles-of-arrival (AoAs). As a result, the sensing echo of the $i$-th target can be extracted from $\tb{r}(t)$ with a receive beamformer $\tb{b}_i=\tb{a}_{N_r}(\tilde{\theta}_{i})$}, expressed as
\begin{align}\label{echo}
	{r}_i(t) =  \gamma_i \tb{b}^{\rm H}_i \tb{a}_{N_r}(\theta_i) \tb{a}_{N_t}^{\rm H}(\theta_i) \tb{f}_i s_i(t-\eta_i)e^{j2\pi \upsilon_i t} + {w}(t).
\end{align}
\revise{We note that angular parameter $\theta_i$ can also be inferred from the receive beamforming process by comparing the gains obtained for different beam directions.} Hence, \eqref{echo} is simplified as
\begin{align}
	{r}_i(t) = G_a s_i(t-\eta_i)e^{j2\pi \upsilon_i t} + {w}_i(t),
\end{align}
where $G_a$ is the composite antenna array gain. By adopting an ideal receive filter and the OTFS demodulation, we finally arrive at the DD domain input-output relationship,
\begin{align}\label{sensing}
R^{i}_{\rm DD}\left[ {l,k} \right]  =& G_a \sum_{k'=0}^{N-1}\sum_{l'=0}^{M-1} H^i_{\rm DD}[l',k']\\&\cdot X^i_{\rm DD}\left[ {(l-l')_M,(k-k')_N} \right] +W^i_{\rm DD}\left[ {l,k} \right],	\nonumber
\end{align}
where $H^i_{\rm DD}[l',k']$ denotes the gain of the $i$-th target at the DD grid (bin) with indices $l',~k'$ corresponding to the delay of $\frac{l'}{M\Delta f}$ and Doppler of $\frac{k'}{N T}$. By stacking all received samples $R^{i}_{\rm DD}$ in a vector $\tb{r}^{i}_{\rm DD}\left[ {l,k} \right]$, \eqref{sensing} is rewritten in a matrix form as
\begin{align}
	\tb{r}^{i}_{\rm DD} = G_a \tb{X}^{i}_{\rm DD} \tb{h}^i_{\rm DD} + \tb{w}^i_{\rm DD},
\end{align}
where the $(Nk'+l')$-th element in $\tb{h}^i_{\rm DD}$ is $H^i_{\rm DD}[l',k']$.
Based on the estimate of sensing channel $\tb{h}^i_{\rm DD}$, it is able to extract the delay and Doppler associated to the $i$-th target.

\section{Preliminary Results and Discussions}

\subsection{Sensing Parameters Estimation}
The classic matched filtering scheme in radar community can be used for obtaining the estimate of $\tb{h}^i_{\rm DD}$, given by
\begin{align}\label{matched_filter}
	\tilde{\tb{r}}^{i}_{\rm DD} = \tb{X}^{i,{\rm H}}_{\rm DD} \tb{r}^{i}_{\rm DD} = G_a \tb{X}^{i,{\rm H}}_{\rm DD}  \tb{X}^{i}_{\rm DD} \tb{h}^i_{\rm DD} + \tb{X}^{i,{\rm H}}_{\rm DD}  \tb{w}^i_{\rm DD}.
\end{align}
Obviously, \eqref{matched_filter} is a linear model and it is straightforward to find an estimator, e.g., maximum likelihood (ML) estimator or linear minimum mean square error (LMMSE) estimator, to obtain the estimate $\hat{\tb{h}}^i_{\rm DD}$. Such estimators are Cram\'{e}r-Rao bound (CRB) achieving if the elements of $\tb{h}^i_{\rm DD}$ are also Gaussian distributed. However, $\tb{h}^i_{\rm DD}$ corresponds to the round-trip delay and Doppler shifts of the $i$-th target, indicating that $\tb{h}^i_{\rm DD}$ is a sparse vector. The delay and Doppler indices can be determined by finding element in $\hat{\tb{h}}^i_{\rm DD}$ that has the largest response, i.e., $[\hat{l}_i, \hat{k}_i]$,
yielding the delay and Doppler estimates $\hat{\eta}_i$ and $\hat{\phi}_i$, respectively\footnote{\revise{In addition to the adopted estimator, the sensing accuracy also depends on the resolutions of the delays and Doppler shifts, which are associated with the bandwidth and signal duration. Therefore, the OTFS frame size, i.e., $M$ and $N$ will also affect the estimation performance.}}. Due to the existence of noise, the above estimation is biased in the low signal-to-noise ratio (SNR) regime, which has been reported in \cite{9109735}.

\subsection{Sensing-assisted OTFS Communications}
From the estimates of delay $\hat{\eta}_i$, Doppler $\hat{\phi}_i$, and angle $\hat{\theta}_i$, we are now able to determine the location as well as the speed of target $i$. Assuming that the BS is located at the origin $[0,0]^{\rm T}$, the location of the $i$-th target $\tb{p}_i = [p_{x,i},p_{y,i}]^{\rm T}$ is written as
\begin{align}
	\hat{p}_{x,i} = \frac{c\hat{\eta}_i\sin\hat{\theta}_i}{2},\\
	{\hat{p}_{y,i} = \frac{c\hat{\eta}_i\cos\hat{\theta}_i}{2}.}
\end{align}
Similarly, the velocity of the $i$-th target $\hat{\tb{v}}_{i}=[\hat{v}_{x,i},\hat{v}_{y,i}]^{\rm T}$ can be obtained based on $\hat{\phi}_i$ and $\hat{\theta}_i$.

According to the definition of OTFS channel model, the Doppler $\nu_i$ is in theory half of the Doppler associated with the sensing echo $\phi_i$. The slow time-varying property of the OTFS communication channel motivates us to reuse the sensing parameters for communications purpose. Relying on the estimates of target locations as well as the speeds, the locations of the targets in the following time instant are predictable. In particular, the predicted location of the $i$-th target can be expressed as
\begin{align}
	\tilde{\tb{p}}_{i} = \hat{\tb{p}}_i + \hat{\tb{v}}_{i} \Delta T,
\end{align}
where $\Delta T$ is the duration of one transmission block. For the speeds of the targets, it is naturally to set the predicted velocity $\tilde{\tb{v}}_i$ identical to the estimated velocity, since the target speed will not have a burst change in a relatively short duration. The predicted locations of targets enable us to further obtain the prediction of the angular parameter $\theta_i$ in the following time instant, based on the geometric relationship as
\begin{align}
	\tilde{\theta}_i =  \arctan \frac{\tilde{p}_{y,i}}{\tilde{p}_{x,i}}.
\end{align}
The predicted angle $\tilde{\theta}_i$ can be used for steering the transmit and receive antennas before transmitting the multi-beam OTFS-ISAC signal and receiving the echoes. Compared to the conventional beam pairing and alignment schemes, using the predicted angle avoids the adoption of dedicated pilots for estimating the angular parameters, which can improve the resource efficiency. The predicted angle can also be contained in the signal to targets, which is used for designing the beamformer $\tb{u}_i$ \cite{9171304,9246715}.

In addition, the prediction of the range distance is obtained as $\tilde{d}_i = \|\tilde{\tb{p}}_{i} \|$. Consequently, the downlink channel gain $h_i$ can be predicted by replacing ${d}_i$ by $\tilde{d}_i$. Moreover, the delay and Doppler indices in the downlink OTFS communications signal model are determined as well by exploiting the predicted range and speed. Motivated by the fact that all communication channel parameters can be predicted at the BS side, it is capable of compensating the channel effects before downlink transmission. Provided sufficiently accurate prediction of the communication channel and compensation at the BS, the received OTFS signal is simplified as
\begin{align}\label{pre_equal}
	Y^i_{\rm DD}\left[ {l,k} \right]  = G_t X^i_{\rm DD} \left[ {l,k} \right] + Z^i_{\rm DD}\left[ {l,k} \right],
\end{align}
with $G_t$ being the composite array gain. It is observed from \eqref{pre_equal} that the communications receiver can perform data detection while bypassing the channel estimation.

\begin{figure}[!t]
\centering
\includegraphics[width=.43\textwidth]{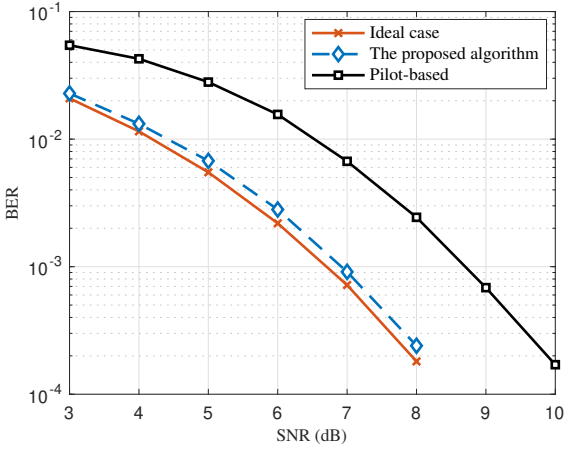}
\caption{{The BER performance of different algorithms for uplink transmission.}}%
\label{fig5}%
\vspace{-1mm}
\end{figure}

To verify the effectiveness of the proposed scheme, we show the bit-error-rate (BER) performance versus the SNR based on the proposed scheme in Fig. \ref{fig5}. The size of OTFS frame is $M = 128$ and $N = 20$. \revise{The BS is operating at a carrier frequency of $3$ GHz and the subcarrier spacing is $6$ kHz. Therefore, the occupied bandwidth is $768$ kHz. The radar cross section (RCS) of the target is assumed to be $25$, given the reflection area of the target being $1\textrm{m}^2$. Finally, the transmit power is set to $40$ dBm.} Moreover, we consider the binary phase shift keying (BPSK) symbol mapping. The speed of the target is randomly generated from the uniform distribution $[10,15]$ m/s. For comparison purpose, the BER performance corresponding to the classic pilot-based beam alignment and channel estimation scheme, and the ideal case with perfect knowledge of channel are illustrated. We can see that our proposed algorithm can approach the performance of the ideal case, which implies the highly accurate channel prediction. \revise{Nevertheless, due to the random noise, the predicted channel may deviate from the actual channel, resulting in a small BER performance gap.} Compared to the classic pilot-based scheme, the proposed algorithm yields a significant BER performance gain. This is because the transmitted signal is known to the BS, providing a high SNR gain to improve the estimation performance of the sensing parameters. Despite the performance improvement, the proposed algorithm also reduces the communication overhead as well as latency of the classic scheme, showing the advantage of adopting OTFS-ISAC signal.

\subsection{Discussions of OTFS-ISAC system}
So far, we have proposed an example of sensing-assisted communications protocol for OTFS-ISAC systems. We show that benefiting from the slow-varying OTFS channel, the parameters extracted from sensing echoes can be exploited for predicting the downlink DD domain communication channel. As a result, no pilots are required either for beam alignment or for channel estimation, leading to a reduced communication overhead. Moreover, since the angle feedback and channel estimation are bypassed, the communications receiver is capable of directly detecting the data symbols, which is attractive for supporting ultra low latency communications in the future networks. Relying on the same DD domain for both communications and sensing, OTFS offers more promising benefits for ISAC systems, which will be briefly discussed in the following.

\subsubsection{Unified Signaling Design Framework}
Let us consider a general single antenna case. When transmitting a vectorized OTFS signal $\tb{x}_{\rm DD}$, after OTFS demodulation, the acquired sensing echo and the received OTFS signal can be written in concise forms as
\begin{align}
\tb{r}_{\rm DD} &= \tb{H}_{s,{\rm DD}} \tb{x}_{\rm DD} + \tb{w}_{\rm DD},\\
\tb{y}_{\rm DD} &= \tb{H}_{c,{\rm DD}} \tb{x}_{\rm DD} + \tb{z}_{\rm DD},
\end{align}
where $\tb{H}_{s,{\rm DD}}$ and $\tb{H}_{c,{\rm DD}} $ denotes the sensing and communication channels in DD domain, respectively. \revise{For sensing, the knowledge of $\tb{x}_{\rm DD}$ at the BS is exploited for inferring the sensing parameters $\bm{\eta}$ from channel $\tb{H}_{s,{\rm DD}}$. In contrast, for communications, the transmitted data symbols $\tb{x}_{\rm DD}$ are detected.} The sensing performance is characterized by ambiguity function and the CRB while the communication performance can be evaluated using channel capacity. \revise{
In particular, the CRB of the sensing parameters $\bm{\eta}$ is derived as \cite{kay1993fundamentals},
\begin{align}
	\textrm{CRB}_{\bm{\eta}} = \textrm{Tr}\left[\mathbb{E}\left[-\frac{\partial^2 \ln p(\mathbf{r}_{\rm DD}|\bm{\eta})}{\partial \bm{\eta}^2}\right]^{-1}\right],
\end{align}
where $p(\mathbf{r}_{\rm DD}|\bm{\eta})$ is the likelihood function, and the capacity of the communication channel $C_{\tb{H}_c}$ can be expressed as
\begin{align}
C_{\tb{H}_c} = \log_2 \textrm{det}\left(\tb{I}_{MN} + \frac{1}{N_0} \tb{H}_{c,\rm{DD}} \tb{R}_x \tb{H}_{c,\rm{DD}}^{\rm H}\right),
\end{align}
where $\tb{R}_x$ is the covariance matrix of the transmitted signal.} The OTFS waveform enables us to formulate the signaling design problem in the same DD domain for achieving satisfactory sensing and communications performance. For example, in a communication-centric scenario, we aim for maximizing the channel capacity $C_{H_c}$ while maintaining the sensing performance. Then, it is straightforward to formulate the following optimization problem
\begin{align}\label{signal_design}
&\max_{\tb{x}_{\rm DD}} C_{\tb{H}_c}\\
\mbox{s.t.}\;\;
\mbox{{C1}}:\; &\textrm{CRB}_{\bm{\eta}} \leq T_{\rm CRB} \notag\\
\mbox{{C2}}:\; &\frac{1}{MN}\|\tb{x}_{\rm DD}\| = P_T, \notag
\end{align}
where C1 indicates the acceptable sensing performance and C2 constraints the power consumption. \revise{A possible solution of \eqref{signal_design} requires a specific waveform design, which can achieve the optimal tradeoff between sensing and communication via taking into account their respectively optimal waveforms in the DD domain.} Moreover, the objective of \eqref{signal_design} can be extended to include both communication and sensing metrics for optimizing OTFS signal that can strike a good trade-off between both functionalities. Instead of directly designing $\tb{x}_{\rm DD}$, we can also design the DD domain precoding matrix $\tb{P}_{\rm DD}$ through optimization given certain constraints such that the transmitted OTFS signal becomes $\tb{P}_{\rm DD}\tb{x}_{\rm DD}$

\subsubsection{Sensing Channel Exploitation}
As discussed in Section III, the delay and Doppler parameters associated with the sensing channel are reused for OTFS communication channel prediction. In this case, we assume that all targets of interest are communication users as well. However, this is not always the case in other scenarios, e.g., only part of the targets in the area of coverage are UEs.
Therefore, a fundamental question is how much information we can glean from the sensing channel $\tb{H}_{s,{\rm DD}}$ for inferring $\tb{H}_{c,{\rm DD}}$. As the first step, we need to characterize the similarity between OTFS communication and sensing channels. A nature choice is to use the cross-correlation matrix $\tb{R}_{cs} = \mathbb{E}\left[\tb{H}_{c,{\rm DD}}\tb{H}_{s,{\rm DD}}^{\rm H}\right]$ and the entries of $\tb{R}_{cs}$ will provide the correlation between sensing targets and communication reflectors corresponding to different paths. In particular, when sensing and communication channels are highly correlated, the matrix $\tb{R}_{cs}$ will be diagonally dominant. Although OTFS modulation offers the opportunity to connect two channels in the same DD domain, new analysis tools and framework are still to be developed to further study the exploitation of sensing channel in OTFS-ISAC system.

\subsubsection{Communications-assisted Sensing}
The OTFS channel model reveals the underlying geometry of the wireless propagation environments. After channel estimation, the estimated delay and Doppler indices can be converted to range distance and speed. Given a temporal sequence of estimated delays corresponding to the $i$-th path, i.e., $\bm{\tau}_i = [{\hat{\tau}_{i,1},...\hat{\tau}_{i,t}}]^{\rm T}$, the location of the $i$-th reflector can be obtained via an ML estimator, i.e.,
\begin{align}
\hat{\tb{p}}_i = \arg\max_{{\tb{p}}_i } p(\bm{\tau}_i|\tb{p}_i),
\end{align}
where $p(\bm{\tau}_i|\tb{p}_i)$ is the likelihood function of measured delays conditioned on the location of the $i$-th reflector. If the reflector is also a moving target, Bayesian estimator such as extended Kalman filtering (EKF) can be used to exploit the state transition information. By combining the sensing information gleaned from the reflected echoes at BS and from the OTFS channel estimation, we can obtain better knowledge of the environment.

\section{Potential Applications of OTFS}

In addition to unifying communications and sensing in the same DD domain, OTFS is also attractive for several other emerging applications, due to its advantages of low complexity implementation, sparse channel representation, and high resilience to large delay and Doppler spread.

\subsection{Underwater Communications}
Due to the propagation medium in underwater communications, mechanical waves, e.g., acoustic wave and seismic wave instead of the radio signal, are usually used for information transmission. The relatively low signal propagation speed of mechanical waves results in large delay and Doppler spreads for underwater communication channels. In general, the delay could spread up to hundreds milliseconds, which is much higher than typical wireless radio channels having the delay level of $\mu$s. The Doppler spread is also severe due to internal waves and the movements of communication transceivers. Moreover, multiple reflectors from the sea-surface and sea floor lead to complex multi-path propagation environments. On the other hand, the inherent wideband propagation in underwater communications will introduce Doppler-scale effect instead of a simple Doppler shift. To characterize different delays and Doppler scales, the multiscale multilag (MSML) channel model is usually used, expressed as
\begin{align}
h(t,\tau) = \sum_{i=1}^P h_i \delta(\tau-\tau_i -\alpha_i t) e^{-j\alpha_i f_c t},
\end{align}
where $P$ is the number of paths, $\alpha_p$ is the time scaling parameter. In fact, in underwater environments, the received signal is a time-delayed, time-scaled, and phase-rotated version of the transmitted signal caused by moving scatterings. Although OFDM has been applied in underwater communications because of its robustness against long delay spread, the inter-carrier interference (ICI) induced by multiple Dopplers will deeply degrade the performance. Compared with OFDM modulation, OTFS modulation could benefit from the possible sparsity and stability of the DD domain channel response, enabling efficient channel estimation and equalization.

\subsection{Optical Transmission}
The optical spectrum offers huge and unregulated bandwidth without electromagnetic interference. Information symbols are transmitted via modulating the luminous intensity of light-emitting diodes (LED) while the received signal is converted to electrical signal using photodiodes. In optical transmission, the channel is generally modeled as static multi-path one. While OTFS is recognized for its superiority in high Doppler scenarios, its capability of exploiting full time-frequency diversity will guarantee its performance in static channels as well. Recent results have shown that OTFS-based optical transmission system achieves a better performance than the conventional OFDM-based counterpart with a single LED \cite{sharma2020performance}. It is worth to extend the research to multi-LED and multi-photodiodes systems based on OTFS modulation.

\subsection{Space-air-ground Integrated Networks (SAGIN)}
To provide ubiquitous connectivity and global coverage, the future communications will be SAGIN. The low-earth-orbit (LEO) satellite communications, the aerial communications, and the vehicular communications will extend the coverage range of the current deployed terrestrial networks. On the one hand, extremely high-mobility is the principal characteristic for the satellite and space communications. Such a high speed will inevitably introduce a very high Doppler shift, which is challenging for reliable information transmission. On the other hand, the networking with different communication devices will result in complex signal transmission scenarios. For example, in high-speed train communications, the environments will change frequently, such as viaducts, cuttings, and tunnels, which imposes great difficulty for robust communications. Relying on the quasi-periodic property, OTFS modulation can support mobile users having a very wide range of speeds. Moreover, it is highly symmetrical to any channel distortions caused by delay and Doppler spreads, making OTFS a good solution for realizing the SAGIN.
Some recent contributions have applied OTFS in LEO system and have verified its effectiveness \cite{shi2022outage}. However, how to design OTFS parameters, such as delay and Doppler resolutions, requires extensive channel measurements and investigations.

%

\section{Conclusions}
In the last part of this tutorial on OTFS modulation, we aim for unveiling its great potentials of supporting future networks. We first studied the OTFS-based ISAC system, which shows the superiority of sensing-assisted OTFS communications. Then several possible benefits of adopting OTFS as the ISAC signaling waveform were discussed. As a further step, we summarized several OTFS modulation-related applications including underwater communications, optical transmission, and SAIGN, which are interesting and of great importance, but have not been completely investigated.

\vspace{-5mm}
\bibliographystyle{IEEEtran}
\bibliography{OTFS_references}

\end{document}